\documentclass{article}

\usepackage{doi}
\usepackage{fullpage}
\usepackage{graphicx}
\usepackage{xspace}
\usepackage{amsmath,amssymb}
\usepackage{hyperref}
\usepackage{booktabs}
\usepackage{xcolor}
\usepackage{olo}

\newcommand{\code}[1]{\texttt{#1}}

\newcommand\myalgo{\textsf{MACER}\xspace}
\newcommand\tracer{\textsf{TRACER}\xspace}
\newcommand\tegcer{\textsf{TEGCER}\xspace}
\newcommand\rlassist{\textsf{RLAssist}\xspace}
\newcommand\deepfix{\textsf{DeepFix}\xspace}

\renewcommand\S{Sec}

\colorlet{mygreen}{green!75!black}
\newcommand{\bbrac}[1]{{[}{#1}{]}}
\newcommand{\rep}[3]{\bbrac{{#1} \bbrac{\textcolor{red}{\texttt{#2}}} \bbrac{\textcolor{mygreen}{\texttt{#3}}}}}
\newcommand{\bigram}[1]{$\langle$\texttt{{#1}}$\rangle$}

\newcommand{\kac}{\ensuremath{\mathsf{k}}\xspace}
\newcommand{\tree}{\ensuremath{\mathsf{tree}}\xspace}
\newcommand{\prot}{\ensuremath{\mathsf{prot}}\xspace}

\graphicspath{{"figures/"}}

\def\mytitle{MACER: A Modular Framework for Accelerated Compilation Error Repair}\relax

\begin{document}

\title{\mytitle}

\author{Darshak Chhatbar$^\ast$ \and Umair Z. Ahmed$^\dagger$ \and Purushottam Kar$^\ast$\\$^\ast$Indian Institute of Technology Kanpur\\$^\dagger$National University of Singapore\\\texttt{\{darshak,purushot\}@cse.iitk.ac.in, umair@comp.nus.edu.sg}}

\date{\today}

\maketitle

\begin{abstract}
Automated compilation error repair, the problem of suggesting fixes to buggy programs that fail to compile, has generated significant interest in recent years. Apart from being a tool of general convenience, automated code repair has significant pedagogical applications for novice programmers who find compiler error messages cryptic and unhelpful. Existing approaches largely solve this problem using a blackbox-application of a heavy-duty generative learning technique, such as sequence-to-sequence prediction (\tracer) or reinforcement learning (\rlassist). Although convenient, such black-box application of learning techniques makes existing approaches bulky in terms of training time, as well as inefficient at targeting specific error types.

We present \myalgo, a novel technique for accelerated error repair based on a modular segregation of the repair process into repair identification and repair application. \myalgo uses powerful yet inexpensive discriminative learning techniques such as multi-label classifiers and rankers to first identify the type of repair required and then apply the suggested repair.

Experiments indicate that the fine-grained approach adopted by \myalgo offers not only superior error correction, but also much faster training and prediction. On a benchmark dataset of 4K buggy programs collected from actual student submissions, \myalgo outperforms existing methods by 20\% at suggesting fixes for popular errors that exactly match the fix desired by the student. \myalgo is also competitive or better than existing methods at all error types -- whether popular or rare. \myalgo offers a training time speedup of $2\times$ over \tracer and $800\times$ over \rlassist, and a test time speedup of $2-4\times$ over both.
\end{abstract}

\section{Introduction}
\label{sec:intro}

The ability to code is a staple requirement in science and engineering and programmers rely heavily on feedback from the programming environment, such as the compiler, linting tools, etc., to correct their programs. However, given the formal nature of these tools, it is difficult to master their effective use without extended periods of exposure. 

\begin{figure}[t]
\includegraphics[width=\columnwidth]{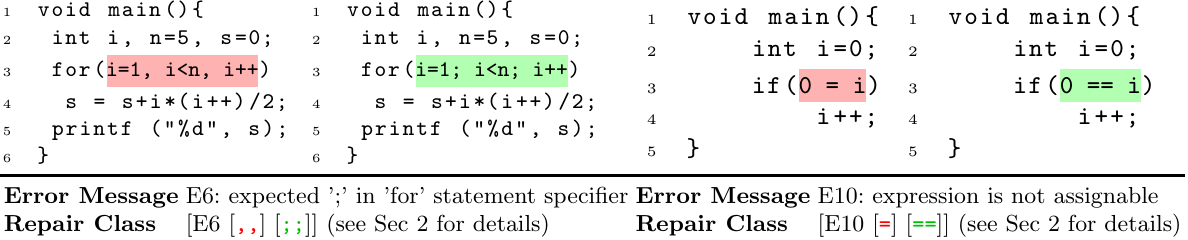}%
\caption{Two examples of actual repairs carried out by \myalgo. The erroneous line in the first example requires multiple replacements to repair the error. Specifically, two occurrences of ',' need to be replaced with ';' as indicated by the repair class description. The erroneous line the second example uses incorrect syntax to check for equality and requires replacing the '=' symbol with the '==' symbol.}
\label{fig:intro-example}%
\end{figure}

Thus, and especially for beginners, these tools can pose a pedagogical hurdle. This is particularly true of compiler error messages which, although always formally correct, can often be unhelpful in guiding the novice programmer on how to correct their error \cite{McCauleyFLMSTZ2008}. This is sometimes due to the terse language used in error messages. For example, see Fig~\ref{fig:intro-example} where the error message uses terms such as ``specifier'' and ''statement'' which may be unfamiliar to a novice. 

At other times this is due to the compiler being unable to comprehend the intent of the user. For example, statements such as \texttt{0 = i;} (where \texttt i is an integer variable) in the C programming language generate an error informing the programmer that the ``expression is not assignable'' (such as with the popular LLVM compiler \cite{lattner2004llvm}). The issue here is merely the direction of assignment but the compiler brings in concepts of expressions and assignability which may confuse a beginner. At still other times, there are compiler quirks, such as reporting a missing terminal semicolon as begin present on the next line.

Several of these issues can be irritating, although not debilitating, to more experienced programmers as well. However, for novices, this frequently means seeking guidance from a human mentor who can then explain the program repair steps in more accessible terms. Given that educational institutions struggle to keep up with increasing student strengths, human mentorship is not a scalable solution~\cite{camp2015booming}.

Consequently, automated program repair has generated a lot of interest in recent years due to its promising applications to programming education and training. A tool that can automatically take a program with compilation errors and suggest repairs to remove those errors can greatly facilitate programming instructors, apart from being a source of convenience for even seasoned programmers.

In this work we report \myalgo, a tool for accelerated repair of programs that fail to compile.

\section{Related Works}
\label{sec:related}
The area of automated program repair has seen much interest in recent years. The \deepfix method \cite{GuptaPKS2017} was one of the first major efforts at using deep learning techniques such as sequence-to-sequence models to locate as well as repair errors. The \tracer method \cite{AhmedKKKG2018} proposed segregating this pipeline into repair line localization and repair prediction and reported better repair performance. This work also introduced the use of the $\pred@\kac$ metric to compilation error repair which demands not just elimination of compilation errors but actually an exact match with the fix desired by the student. This was a much more stringent metric than the prevailing \emph{repair accuracy} metric which simply counted reductions in compilation errors.

Recent works have focused on several other aspects of this problem. The \rlassist method \cite{GuptaKS2019} introduced self learning techniques using reinforcement learning to eliminate the need for training data. However, the technique offers slow training times. The work of \cite{HajipourBF2019} proposes to use generative techniques using variational auto-encoders to introduce diversity in the fixes suggested by the technique. The work of \cite{VasicKMBS2019} focuses on locating and repairing a special class of errors called \emph{variable-misuse} errors which are logical errors where programmers use an inappropriate identifier, possibly due to confusion in identifier names. The \tegcer method \cite{AhmedSSK2019} focuses not on error repair but rather \emph{repair demonstration} by showing students, fixes made by other students on similar errors which can be argued to have greater pedagogical utility. 

The works \deepfix, \rlassist and \tracer most directly relate to our work and we will be comparing to all of them experimentally. \myalgo outperforms all these methods in terms of repair accuracy, exact match ($\pred@\kac$) accuracy, training and prediction time, or all of the above.

\section{Our Contributions}
\label{sec:contributions}
\myalgo makes the following key contributions to compilation error repair
\begin{enumerate}
	\item \myalgo sets up a modular pipeline that, in addition to locating lines that need repair, further segregates the repair pipeline by identifying \emph{what} is the type of repair needed on each line (the \emph{repair-class} of that line), and \emph{where} in that line to apply that repair (the \emph{repair-profile} of that line). This presents a significant departure from previous works like \tracer and \deepfix that rely on a heavy-duty generative mechanism to perform the last two operations (repair type identification and application) in a single step to directly generate repair suggestions.
	\item Although convenient, these generative mechanisms used in previous works come at a cost -- not only are they expensive at training and prediction, but their one-step approach also makes it challenging to fine tune their method to focus more on certain types of errors than others. We show that \myalgo on the other hand, is able to specifically target certain error types. Specifically, \myalgo is able to pay individual attention to each repair class to offer superior error repair.
	\item \myalgo introduces techniques used in large-scale multi-class and multi-label learning tasks, such as hierarchical classification and reranking techniques, to the problem of program repair. To the best of our knowledge, these highly efficient and scalable techniques have hitherto not been applied to the problem of compilation error repair.
	\item \myalgo accurately predicts the repair class (see Tab~\ref{tab:reranking-benefits}). Thus, instructors can manually rewrite helpful feedback (to accompany \myalgo's suggested repair) for popular repair classes which may offer greater pedagogical value.
	\item We present a highly optimized implementation of an end-to-end tool-chain\footnote{The \myalgo tool-chain is available at https://github.com/purushottamkar/macer/} for compilation error repair that effectively uses these scalable techniques. \myalgo's repair pipeline is end-to-end and entirely automated i.e. steps such as creation of repair classes can be replicated for any programming language for which static type inference is possible.
	\item The resulting implementation of \myalgo not only outperforms existing techniques on various metrics, but also offers training and prediction times that are several times to orders of magnitude faster than those of existing techniques.
\end{enumerate}

\section{Problem Setting and Data Preprocessing}
\label{sec:formulation}
\myalgo learns error repair strategies given training data in the form of several pairs of programs, with one program in the pair failing to compile (called the \emph{source program}) and the other program in the pair being free of compilation errors (called the \emph{target program}). Such a \emph{supervised} setting is standard in previous works in compilation error repair \cite{AhmedKKKG2018,GuptaKS2019}. Similar to \cite{AhmedKKKG2018}, we train only on those pairs where the two programs differ in a single line (e.g. in Fig~\ref{fig:intro-example}, the programs differ only in line 3).

However, we stress that \myalgo is able to perform repair on programs where multiple lines may require repairs as well, and we do include such datasets in our experiments. The differing line in the source (resp. target) program is called the \emph{source line} (resp. \emph{target line}). With every such program pair, we also receive the errorID and message generated by the Clang compiler \cite{lattner2004llvm} when compiling the source program. Tab~\ref{tab:comp-errorID} lists a few errorIDs and error messages. It is clear from the table data that some error types are extremely rarely encountered whereas others are very common.

\begin{table}%
\centering
\caption{{Some examples of the 148 compiler errorIDs listed in decreasing order of their frequency of occurrence in the data (reported in the \emph{Count} column). It is clear that some error types are extremely frequent whereas other error types rarely occur in data. The symbol $\Box$ is a placeholder for program specific tokens such as identifiers, reserved keywords, punctuation marks etc. For example, a specific instance of errorID E6 is shown in Figure~\ref{fig:intro-example}. A specific instance of errorID E1 could be ``Expected \texttt ; after expression''.}}
\label{tab:comp-errorID}
\begin{tabular}{lll}
\toprule
ErrorID & Error Message & Count\\
\midrule
E1 & Expected $\Box$ after expression & 4999\\
E2 & Use of undeclared identifier $\Box$ & 4709\\
E3 & Expected expression $\Box$ & 3818\\
E6 & Expected $\Box$ in $\Box$ statement specifier & 720\\
E10 & Expression is not assignable & 538\\
E23 & Expected ID after return statement & 128\\
E57 & Unknown type name $\Box$ & 23\\
E76 & Non-object type $\Box$ is not assignable & 11\\
E98 & variable has incomplete type ID & 3\\
E148 & Parameter named $\Box$ is missing & 1\\
\bottomrule
\end{tabular}
\end{table}

\subsection{Notation}
We use angular brackets to represent n-grams. For example, the statement $a = b + c;$ contains the unigrams \bigram{a}, \bigram{=}, \bigram{b}, \bigram{+}, \bigram{c} and \bigram{;}, as well as contains the bigrams \bigram{a =}, \bigram{= b}, \bigram{b +}, \bigram{+ c}, \bigram{c ;} and \bigram{; EOL}. When encoding bigrams, we include an end-of-line character \texttt{EOL} as well. This helps \myalgo distinguish this location since several repairs (such as insertion of expression termination symbols) require edits at the end of the line.

\subsection{Feature Encoding}
\label{sec:feat-enc}
The source lines contain several user-defined literals and identifiers (variable names) which can be diverse but are not informative for error repair. To avoid overwhelming the machine learning method with these uninformative tokens, it is common in literature to reduce the input vocabulary size. \myalgo does this by replacing literals and identifiers with their corresponding abstract LLVM token type while retaining keywords and symbols. An exception is string literals where format-specifiers (such as \verb+%d+ and \verb+%s+) are retained as is, since these are often a source of errors themselves.

For example, the raw or \emph{concrete} statement \code{int abc = 0;} is converted into the \emph{abstract} statement \code{int VARIABLE\_INT = LITERAL\_INT ;}. An attempt is made to infer the datatypes of identifiers (which is possible even though compilation fails on these programs since the compiler is often nevertheless able to generate a partial symbol table while attempting compilation). Undeclared/unrecognized identifiers are replaced with a generic token \texttt{INVALID}. Such abstraction is common in literature. \deepfix \cite{GuptaPKS2017} replaces each program variable name with a generic identifier \code{ID} and removes the contents of string literals. We use the abstraction module described in \tracer \cite{AhmedKKKG2018} that retains type information, which is helpful in fixing type errors.

For our data, this abstraction process yielded a vocabulary size of 161 tokens and 1930 unique bigrams. Both uni and bigrams were included in the representation since this feature representation will be used to predict the repair class of the line which involves predicting which tokens (i.e. unigrams) need to be replaced/deleted/inserted, as well as be used to predict the repair profile of the line which involves predicting bigrams as locations. Thus, having uni and bigrams natively in the representation eased the task of these classifiers. Including trigrams in the representation did not offer significant improvements but increased training and prediction times.

\myalgo represents each source line as a 2239 dimensional binary vector. The first 148 dimensions in this representation store a one-hot encoding of the compiler errorID generated on that source line (see Tab~\ref{tab:comp-errorID} for examples). The next 161 dimensions store a one-hot unigram feature encoding of the source line and the remaining 1930 dimensions store a one-hot bigram feature encoding of the abstracted source line. We used one-hot encodings rather than TF-IDF encodings since the additional frequency information for uni and bigrams did not offer any predictive advantage. We also found the use of trigrams to not significantly increase performance but make the method slower. It is important to note that the feature creation step does not use the target line in any manner. This is crucial to allow feature creation for test examples as well.

\begin{table}%
\centering
\caption{{Some examples of the 1016 repair classes used by \myalgo listed in decreasing order of their frequency of occurrence (frequencies reported in the column \emph{Count}). For example, Class C2 concerns the use of undeclared identifiers and the solution is to replace the undeclared identifier (\texttt{INVALID} token) with an integer variable or literal. A $\emptyset$ indicates that no token need be inserted/deleted for that class. For example, no token need be inserted to perform repair for repair class C22 whereas no token need be deleted to perform repair for repair class C115. See the text for a description of the notation used in the second column.}}
\label{tab:repair-classID}
\begin{tabular}{llll}
\toprule
Class ID & \rep{ErrorID}{Del}{Ins} & Type & Count\\
\midrule
C1 & \rep{E1}{$\emptyset$}{;} & Insert & 3364\\
C2 & \rep{E2}{INVALID}{INT} & Replace & 585\\
C12 & \rep{E6}{,}{;} & Replace & 173\\
C22 & \rep{E23}{;}{$\emptyset$} & Delete & 89\\
C31 & \rep{E6}{,,}{;;} & Replace & 62\\
C64 & \rep{E3}{)}{$\emptyset$} & Delete & 33\\
C99 & \rep{E45}{==}{=} & Replace & 19\\
C115 & \rep{E3}{$\emptyset$}{`} & Insert & 16\\
C145 & \rep{E24}{.}{->} & Replace & 11\\
C190 & \rep{E6}{for}{while} & Replace & 9\\
\bottomrule
\end{tabular}
\end{table}

\begin{figure}[t]%
	\centering
	\includegraphics[width=0.5\linewidth]{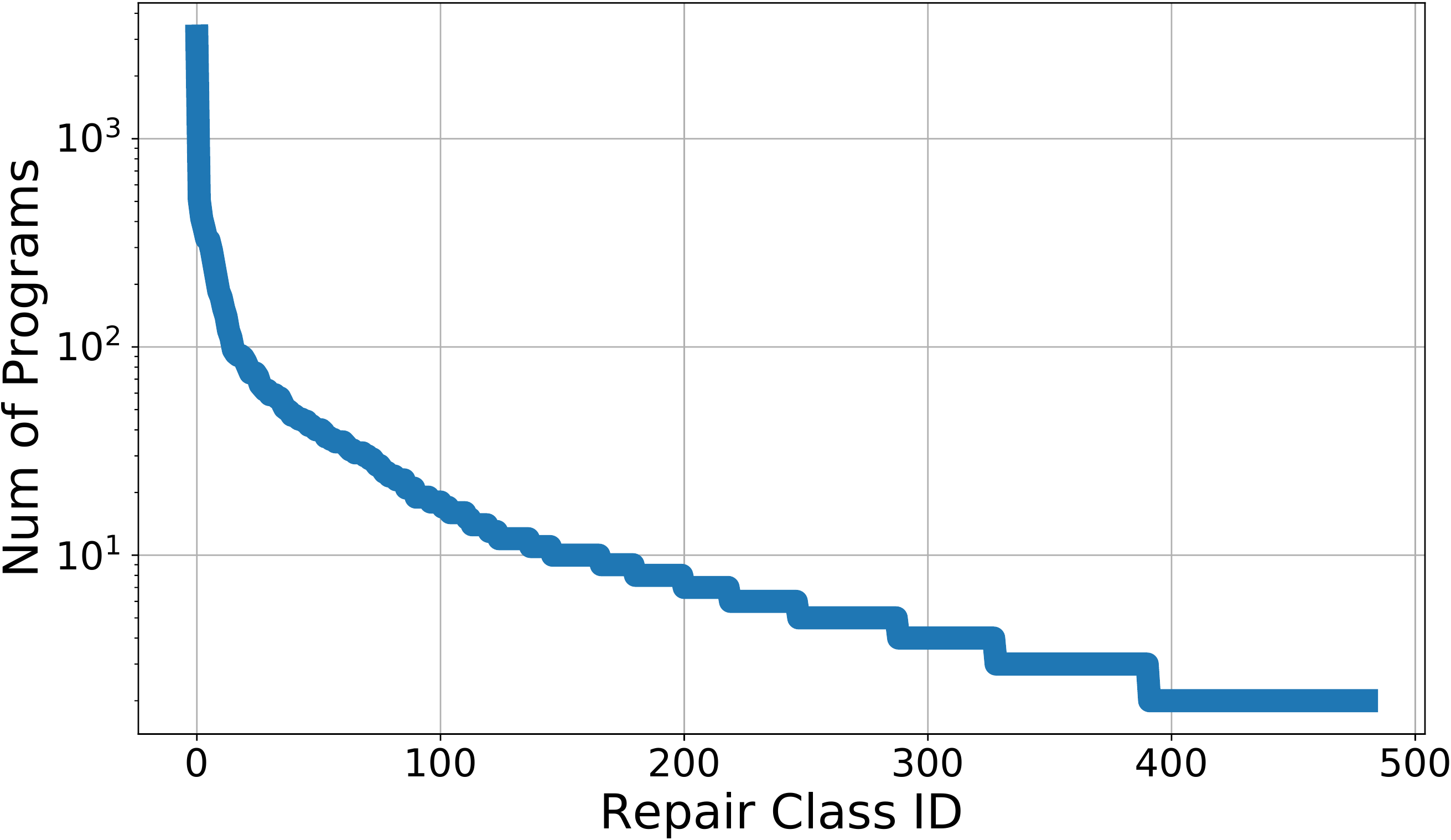}
	\caption{{Repair classes generated by \myalgo arranged in descending order of the number of training programs associated with them. Only the 500 most popular classes are shown. The classes exhibit heavy-tailed behavior: less than 400 of the 1016 classes have 3 or more training data points associated with them. On the other hand, the top 10 classes have more than 200 training points each.}}%
	\label{fig:heavy-tail}%
\end{figure}

\subsection{Repair Class Creation}
\label{sec:repair-class-create}
The \emph{repair class} of a source line encodes \emph{what} repair to apply to that line. As noted in Table~\ref{tab:comp-errorID}, the Clang compiler offers 148 distinct errorIDs. However, diverse repair strategies may be required to handle all instances of any given errorID. For example, errorID E6 in Fig~\ref{fig:intro-example} can of course signal missing semicolons within the header of a \texttt{for} loop as the example indicates, but it can also be used by the compiler to signal missing semicolons \texttt{;} at the end of a \texttt{do-while} block, as well as missing colons \texttt{:} in a \texttt{switch} case block.

To consider the above possibilities, similar to \tegcer \cite{AhmedSSK2019}, we first expand the set of 148 compiler-generated errorIDs into a much bigger set of 1016 \emph{repair classes}. It is notable that these repair classes are generated automatically from training data and do not require any manual supervision. For each training instance, token abstraction (see \S~\ref{sec:feat-enc}) is done on both the source and target lines and a diff is taken between the two. This gives us the set of tokens that must be deleted from the (abstracted) source line, as well as those that must be inserted into the source line, in order to obtain the (abstracted) target line.

A tuple is then created consisting of the compiler errorID for that source line, followed by an enumeration of tokens that must be deleted (in order of their occurrence in the source line from left to right), followed by an enumeration of tokens that must be inserted (in order of their insertion point in the source line from left to right). Such a tuple of the form
\begin{center}
	\rep{ErrID}{TOK$^-_1$ TOK$^-_2$ \ldots}{TOK$^+_1$ TOK$^+_2$ \ldots}
\end{center}
\noindent is called a \emph{repair class}. We identified 1016 such classes. A repair class requiring no insertions (resp. deletions) is called a \emph{Delete} (resp. \emph{Insert}) repair class. A repair class requiring as many insertions as deletions with insertions at exactly the locations of the deletions is called a \emph{Replace} repair class. Tab~\ref{tab:repair-classID} illustrates a few repair classes. Repair classes exhibit a heavy tail (see Fig~\ref{fig:heavy-tail}) with popular classes having hundreds of training points whereas the vast majority of (rare) repair classes have merely single digit training instances.

\subsection{Repair Profile Creation}
\label{sec:repair-profile-create}
The \emph{repair profile} of a source line encodes \emph{where} in that line to apply the repair encoded in its repair class. For every source line, taking the diff of the abstracted source and abstracted target lines (as done in \S~\ref{sec:repair-class-create}) also tells us which bigrams in the abstracted source line require some edit operation (insert/delete/replace) in order to obtain the abstracted target line.

The repair profile for a training pair stores the identity of these bigrams which require modification for that source line. A one-hot representation of the set of these bigrams i.e. a binary vector $\vr \in \bc{0,1}^{1930}$ is taken to be the repair profile of that source line. We note that the repair profile is a sparse fixed-dimensional binary vector (that does not depend on the number of tokens in the source line) and ignores repetition information. Thus, even if a bigram requires multiple edit operations, or even if a bigram appears several times in the source line and only one of those occurrences requires an edit, we record a 1 in the repair profile corresponding to that bigram. This was done in order to simplify prediction of the repair profile for erroneous programs at testing time.

\subsection{Working Dataset}
After the steps in \S~\ref{sec:feat-enc}, \ref{sec:repair-class-create}, and \ref{sec:repair-profile-create} have been carried out, we have with us, corresponding to every source-target pair in the training dataset, a class-label $y^i \in [1016]$ telling us the repair class for that source line, a feature representation $\vx^i \in \bc{0,1}^{2239}$ that tells us the errorID, and the uni/bigram representation of the source line, and a sparse Boolean vector $\vr^i \in \bc{0,1}^{1930}$ that tells us the repair profile. Altogether, this constitutes a dataset of the form $\bc{(\vx^i,y^i,\vr^i)}_{i=1}^n$.

\section{\myalgo: Methodology}
\label{sec:method}

\begin{figure*}[t]%
\centering
\includegraphics[width=\linewidth]{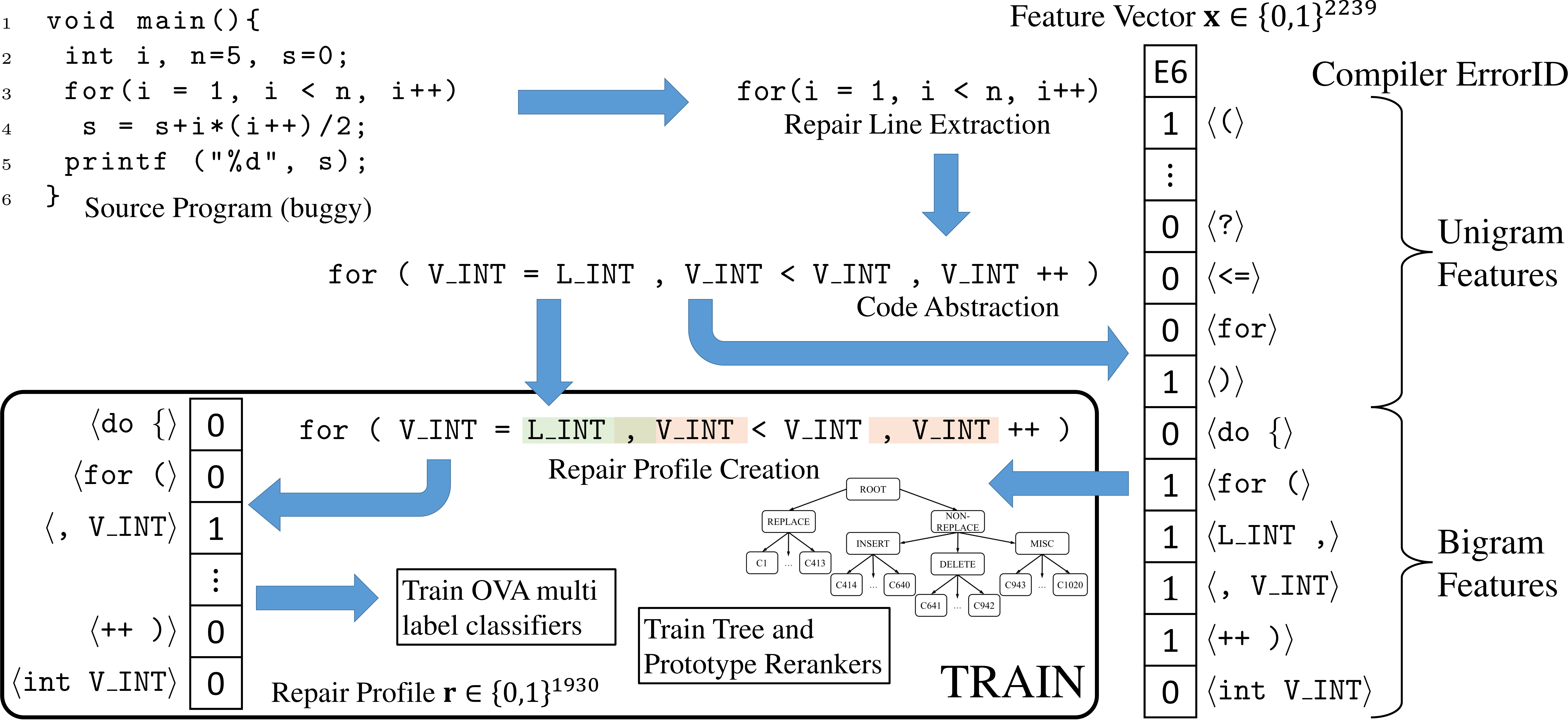}%
\caption{{The training pipeline proposed by \myalgo, illustrated using the example used in Fig~\ref{fig:intro-example}. In the example above, \code{L\_INT} and \code{V\_INT} are shorthand for \code{LITERAL\_INT} and \code{VARIABLE\_INT}.}}%
\label{fig:pipeline_train}%
\end{figure*}

\begin{figure*}[t]%
\centering
\includegraphics[width=0.5\linewidth]{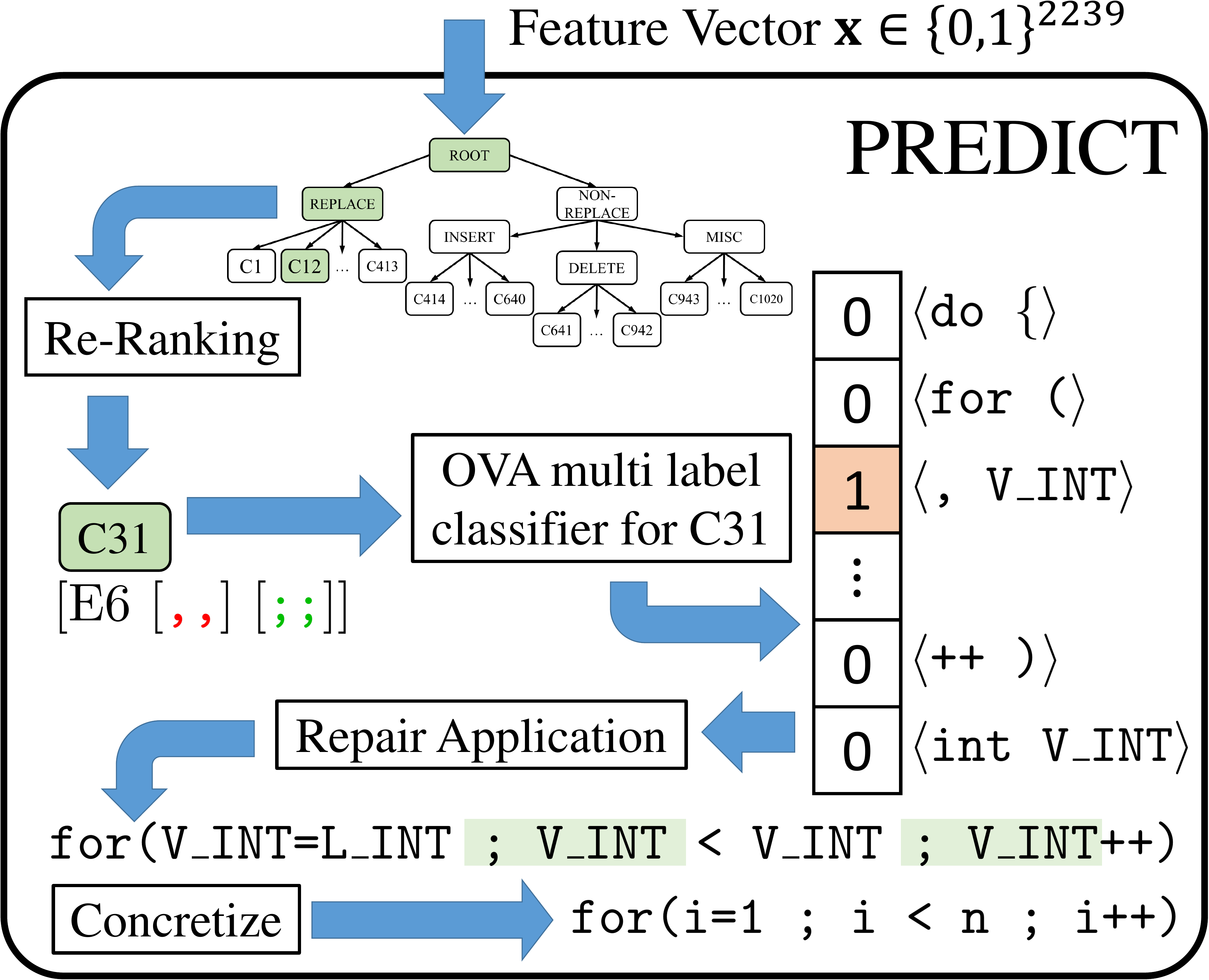}%
\caption{{The repair pipeline proposed by \myalgo, illustrated using the example used in Fig~\ref{fig:intro-example}. To illustrate the benefit of reranking, we depict a situation where a wrong repair class gets highest score from the classification tree, but reranking corrects the error. Tab~\ref{tab:reranking-benefits} shows that this is indeed common. The repair line extraction and code abstraction steps are common to training and prediction.}}%
\label{fig:pipeline_pred}%
\end{figure*}

\myalgo (Modular Accelerated Compilation Error Repair) segregates the repair process into six distinct steps
\begin{enumerate}
	\item \textbf{Repair Lines}: Locate within the source code, which line(s) are erroneous and require repair.
	\item \textbf{Feature Encoding}: For each of the identified lines, perform code abstraction and obtain a 2239-dimensional feature vector (see \S~\ref{sec:feat-enc}).
	\item \textbf{Repair Class Prediction}: Use the feature vector to predict which of the 1016 repair classes is applicable i.e. which type of repair is required.
	\item \textbf{Repair Localization}: Use the feature vector to predict locations within the source lines at which repairs should be applied.
	\item \textbf{Repair Application}: Apply the predicted repairs at the predicted locations
	\item \textbf{Repair Concretization}: Undo code abstraction and compile.
\end{enumerate}
Although previous works do incorporate some of the above steps, e.g., \tracer incorporates code abstraction and locating repair lines within the source code, \myalgo departs most notably from previous approaches in segregating the subsequent repair process into repair class prediction, localization, and application steps. Among other things such as greater training and prediction speed, this allows \myalgo to learn a customized repair location and repair application strategy for different repair classes which can be beneficial. For instance, if it is known that the repair required is the insertion of a semi-colon, then the location where the repair must be performed is narrowed down significantly.

In contrast, existing methods expect a generative mechanism such as sequence-to-sequence prediction or reinforcement learning, to jointly perform all these tasks. This precludes any opportunity to exploit the type of repair required to perform better on specific repair types, apart from making these techniques slow at training and prediction. Below we detail the working of each of the above steps.

\subsection{Repair Lines}
\label{sec:err-line}
One of the key tasks of a compiler is to report line numbers where an error was encountered. However, this does not necessarily correspond to the location where the repair must be performed. In our training data set where errors are localized to a single line, the repair line location was the same as the compiler reported line-number in only about 80\% of the cases.

Existing works have used different techniques for repair line localization. \rlassist \cite{GuptaKS2019} use reinforcement learning to perform localization by navigating the program using movement based actions to maximize a predefined reward. \deepfix \cite{GuptaPKS2017} trains a dedicated neural-network to identify suspicious tokens (and hence their location) to achieve around 86\% repair line localization accuracy. \tracer \cite{AhmedKKKG2018} relies on compiler reported line numbers and considers two additional lines, one above and one below the compiler error line, and obtains a localization accuracy of around 87\%. \myalgo uses this same technique which gave a \emph{repair line localization} recall of around 90\% on our training dataset.

\begin{figure}[t]%
\centering
\includegraphics[width=0.75\columnwidth]{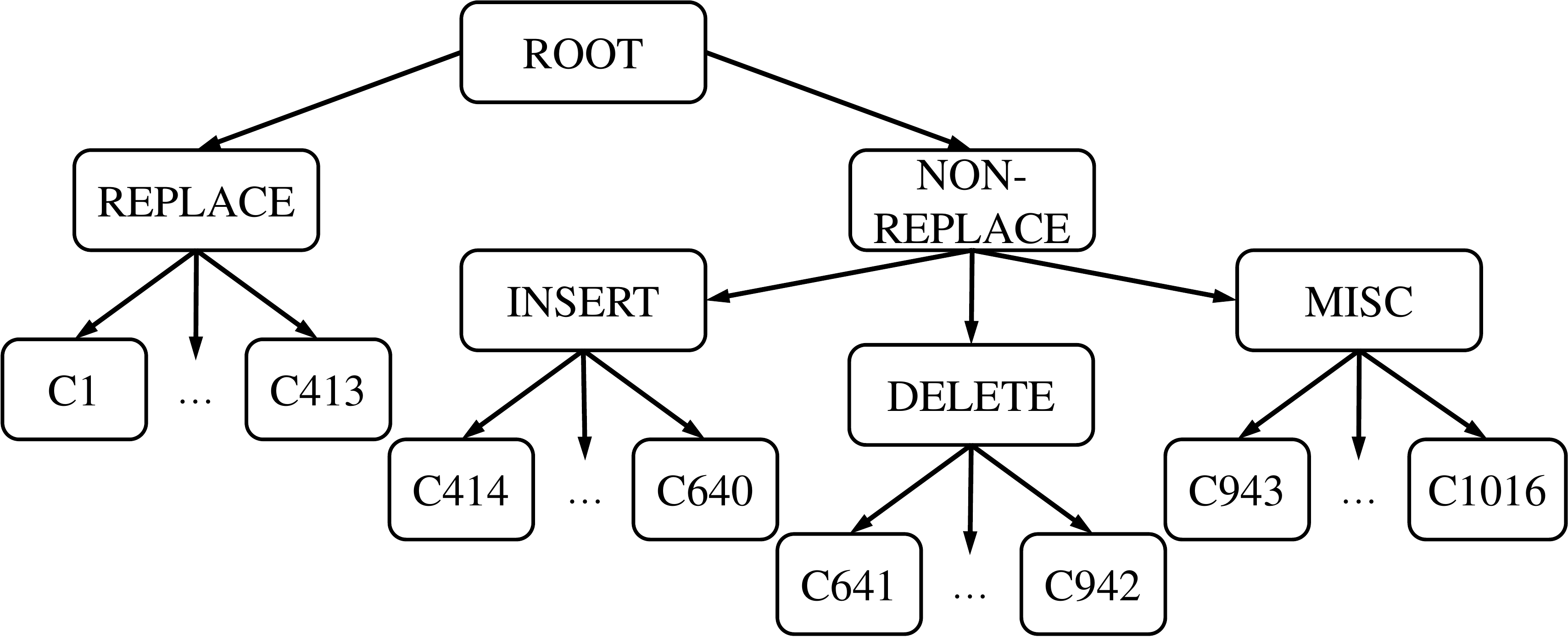}%
\caption{The prediction hierarchy used by \myalgo to predict the repair class.}%
\label{fig:hierarchy}%
\end{figure}

\subsection{Repair Class Prediction}
\label{sec:repair-class-pred}
As outlined in \S~\ref{sec:formulation}, \myalgo considers 1016 repair classes which is quite large. In order to make fast and accurate predictions for the correct repair class that apply to a given source line, \myalgo uses hierarchical classification techniques that are popular in the domain of large-scale multi-class and multi-label classification problems \cite{JasinskaDB-FPKH2016,PrabhuKHAV2018}.

As there exists a natural hierarchy in our problem setting, we found it suitable (as suggested by \cite{JasinskaDB-FPKH2016}) to use a fixed hierarchy rather than a learnt hierarchy. Given that a significant fraction of repair classes (around 40\%) involve replacement repairs, we found it advantageous to first segregate source lines that require replacement repairs from others.

The classification hierarchy used by \myalgo is shown in Fig~\ref{fig:hierarchy}. The root node decides whether a source line requires a replacement or some other form of repair using a feed-forward network with 2 hidden layers with 128 nodes each and trained on cross entropy loss. All other internal nodes use a linear one-vs-rest classifier trained on cross entropy loss to perform their respective multi-way splits.

It is well-known \cite{JainPV2016} that discriminative classifiers struggle to do well on rare classes due to paucity of data. Our repair classes do exhibit significant heavy-tailed behavior (see Fig~\ref{fig:heavy-tail}) with most classes occurring infrequently and only a few being popular. To improve \myalgo's performance, we first augment the classification tree into a ranking tree that ranks classes instead of just predicting one class, and then introduce a reranking step which modifies the ranking given by the ranking tree.

\subsubsection{Repair Class Ranking}
\label{sec:repair-class-rank}
We converted the classification tree into a probabilistic ranking tree that could assign a likelihood score to each repair class. More specifically, given the feature representation of a source line $\vx \in \bc{0,1}^{2239}$, the tree is used to assign, for every repair class $c \in [1016]$, a likelihood score $s^\tree_c(\vx) := \P{y = c \cond \vx}$. We followed a process similar to (\cite{JasinskaDB-FPKH2016,PrabhuKHAV2018}) to obtain these likelihood scores from the tree. This construction is routine and detailed in the appendix \S~\ref{app:repair-class-rank-app}. Although these scores $s^\tree_c(\vx)$ can themselves be used to rank the classes, doing so does not yield the best results. This is due to the large number of extremely rare repair classes (Tab~\ref{tab:repair-classID} shows that only $\approx$ 150 of the 1016 repair classes have more than 10 training examples).

\begin{table}[t]
	\centering
	\small{
	\caption{{Performance benefits of reranking. The table shows the performance accuracy (in terms of various ranking metrics) achieved by \myalgo in predicting the correct repair class. The first three columns report $\topl@k$ i.e. the fraction of test examples on which the correct errorID or correct repair tokens were predicted within the top $k$ locations of the ranking. The last column reports the mean-average precision i.e. the average reciprocal rank at which the correct repair class was predicted in terms of tokens to be inserted or deleted. Note that in all cases, reranking significantly boosts the performance. In particular, the last column indicates that reranking ensures that the correct tokens to be inserted/deleted were almost always predicted within the first two ranks.}}
	\label{tab:reranking-benefits}
	\begin{tabular}{*5c}
		\toprule
		  & $\topl@1$ & $\topl@3$ & $\topl@5$ & MAP\\
		\midrule
		Reranking Off (use $s^\tree_c(\vx)$ to rank repair classes) & 0.66 & 0.83 & 0.87 & 0.40\\
		Reranking On (use $0.8\cdot s^\tree_c(\vx) + 0.2\cdot s^\prot_c(\vx)$ instead) & 0.67 & \textbf{0.88} & \textbf{0.90} & \textbf{0.50}\\
		\bottomrule	
	\end{tabular}
	}
\end{table}

\subsubsection{Repair Class Reranking}
\label{sec:reranking}
To improve classification performance on rare repair classes, \myalgo uses \emph{prototype classifiers} \cite{JainPV2016,PrabhuKGDHAV2018} that have been found to be effective and scalable when dealing with a large number of rare classes. Suppose a repair class $c \in [1016]$ is associated with $n_c$ training points. The k-means algorithm is used to obtain $k_c = \ceil{\frac{n_c}{25}}$ clusters out of these $n_c$ training points and the centroids of these clusters, say $\tilde\vx_c^1,\ldots,\tilde\vx_c^{k_c}$, are taken as \emph{prototypes} for this repair class. This is repeated for all repair classes.

At test time, given a source line $\vx \in \bc{0,1}^{2239}$, these prototypes are used to assign a new score to each repair class as follows
\[
s^\prot_c(\vx) := \max_{k \in [k_c]}\ \exp\br{-\frac12\norm{\vx - \tilde\vx_c^k}_2^2}
\]
Thus, for each repair class, the source line searches for the closest prototype of that repair class and uses it to generate a score.

\myalgo uses the scores assigned by the probabilistic ranking tree and those assigned by the prototypes to get a combined score as $s_c(\vx) = 0.8\cdot s^\tree_c(\vx) + 0.2\cdot s^\prot_c(\vx)$ (the constants 0.8, 0.2 are standard in literature and we did not tune them). \myalgo uses this combined score $s_c(\vx)$ to rank repair classes in decreasing order of their applicability. Tab~\ref{tab:reranking-benefits} outlines how the reranking step significantly boosts \myalgo's ability to accurately predict the relevant compiler errorID and the repair class. \S~\ref{sec:exps} will present additional ablation studies that demonstrate how the reranking step boosts not just the repair class prediction accuracy, but \myalgo's error repair performance as well.

\subsection{Repair Localization}
\label{sec:repair-loc}
Having predicted the repair class, \myalgo proceeds to locate regions within the source line where those repairs must be made. \myalgo reformulates this a problem of predicting which bigram(s) within the source line require edits (multiple locations may require edits on the same line). Note that this is exactly the same as predicting the repair profile vector of that source line (apart from any ambiguity due to the same bigram appearing multiple times in the line which is handled during repair application).

This observation turns repair localization into a multi-label learning problem, that of predicting the sparse Boolean repair profile vector $\vr$ using the source feature representation $\vx$ and the predicted repair class $\hat y$ as inputs. Each of the 1930 bigrams in our vocabulary, now turns into a potential ``label'' which if turned on, indicates that repair is required at bigrams of that type. Fig~\ref{fig:pipeline_pred} explains the process pictorially.

\myalgo adopts the ``one-vs-rest'' (OVR) approach that is a state-of-the-art in large-scale multi-label classification \cite{BabbarS2017}. Thus, 1930 binary classifiers are trained (using standard implementations of decision trees), each one predicting whether that particular bigram needs repair or not. At test time, classifiers corresponding to all bigrams present in the source line are queried as to whether the corresponding bigrams require repair or not.

\myalgo trains a separate OVR multi-label classifier per repair class that is trained only on training points of that class (as opposed to having a single OVR classifier handle examples of all repair classes). Training repair-class specific localizers improved performance since the kind of bigrams that require edits for replacement repairs e.g. substituting \texttt{=} with \texttt{==}, are very different from the kind of bigrams that require edits for insertion repairs e.g. inserting a semicolon \texttt{;}. At test time, after predicting the repair class for a source line, \myalgo invokes the OVR classifier of the predicted repair class to perform repair localization. During repair localization, we only invoke OVR classifiers corresponding to bigrams that are actually present in the source line. Thus, we are guaranteed that any bigrams predicted to require edits will always be present in the source line. In our experiments, we found this strategy to work well with \myalgo offering a Hamming loss of just 1.43 in terms of predicting the repair profile as a Boolean vector. Thus, on an average, only about one bigram was either predicted to require repair when it did not, or not predicted to require repair when it actually did.

\subsection{Repair Application}
\label{sec:repair-app}
The above two steps provide \myalgo with information on \emph{what} repairs need to be applied as well as \emph{where} they need to be applied. Frugal but effective techniques are then used to apply the repairs which we discuss in this subsection. Let $\cB$ denote the ordered set of all bigrams (and their locations, ordered from left to right) in the source line which were flagged by the repair localizer as requiring edits. For example, if the repair localizer predicts the bigram \bigram{, VARIABLE\_INT} to require edits and this bigram appears twice in the source line (note that this would indeed happen in the example in Fig~\ref{fig:intro-example}), then both those bigrams would be included in $\cB$. This is repeated for all bigrams flagged by the repair localizer. Below we discuss the repair application strategy for various repair class types.

\subsubsection{Insertion Repairs}
Recall that these are repairs where no token needs to be deleted from the source line but one or more tokens need to be inserted. We observed that in an overwhelmingly large number of situations that require multiple tokens to be inserted, all tokens need to be inserted at the same location, for instance the repair

\texttt{for(i=0;i<5)} $\rightarrow$ \texttt{for(i=0;i<5;i++)}

\noindent has the repair class \rep{E6}{$\emptyset$}{; VARIABLE\_INT ++} and requires three tokens, a semicolon \texttt{;}, an integer variable identifier, and the increment operator \texttt{++} to be inserted, all at the same location i.e. within the bigram \bigram{5 )} which is abstracted as \bigram{LITERAL\_INT )}.

Thus, for insert repairs, \myalgo concatenates all tokens marked for insertion in the predicted repair class and attempts insertion of this ensemble into all bigrams in the set $\cB$. Attempting insertion into a single bigram itself requires 3 attempts since each bigram offers 3 positions for insertion within itself. After each attempt, \myalgo concretizes the resulting program (see below) and attempts to compile it. \myalgo stops at a successful compilation and keeps trying otherwise.

\subsubsection{Deletion Repairs}
Recall that these are repairs where no token needs to be inserted into the source line but one or more tokens need to be deleted. In this case, \myalgo scans the list of tokens marked for deletion in the predicted repair class from right to left. For every such token, the first bigram in the ordered set $\cB$ (also scanned from right to left) that has that token, gets edited by deleting that token. Once all tokens in the repair class are exhausted, \myalgo concretizes the resulting program (see below) and attempts to compile it.

\subsubsection{Replace Repairs}
Recall that these are repairs where an insertion and a deletion, both happen at the same location and this process may be required multiple times. In such cases, \myalgo scans the list of tokens marked for deletion in the predicted repair class from right to left and also considers the corresponding token marked for insertion. Let this pair be (\texttt{TOK}$^-$, \texttt{TOK}$^+$). As in the deletion repair case, the first bigram in the ordered set $\cB$ (also scanned from right to left) that contains \texttt{TOK}$^-$, gets edited by deleting \texttt{TOK}$^-$ from that bigram and inserting \texttt{TOK}$^+$ in its place. Once all tokens in the repair class are exhausted, \myalgo concretizes the resulting program (see below) and attempts to compile it.

\subsubsection{Miscellaneous Repairs}
In the most general case, an unequal number of tokens may need to be inserted and deleted from the source line, that too possibly at varied locations. Handling all cases in this situation separately is unwieldy and thus, \myalgo adopts a generic approach which works well in a large number of cases. First, \myalgo ignores the insertion tokens in the repair class and performs edits as if the repair class were a deletion type class. Subsequently, it considers the insertion tokens (all deletion tokens having been considered by now) and processes the resulting edited line as if it were an insertion type class.

\subsubsection{Repair Concretization}
The above repair process generates a line that still contains abstract LLVM tokens such as \verb+LITERAL_INT+. In order to make the program compilable, these abstract tokens are replaced with concrete program tokens such as literals and identifiers through an approximate process reverses the abstraction. To do this, we replace each abstract token with the most recently used concrete variable/literal of the same type, that already exists in the current scope.

This is an approximate process since the repair application could suggest the insertion of a particular type of variable, which does not exist in the current scope. For example, if the repair application stage suggests the insertion of a variable of type \code{Variable\_Float}, then at least one floating point variable should be declared in the same scope as the erroneous line. Nevertheless, we observe that this concretization strategy of \myalgo is able to recover the correct replacement in 90+\% of the instances in our datasets.

\myalgo considers each candidate repair line reported by the repair line localizer (recall that these include compiler-reported lines as well as lines immediately above and below those lines). For each such candidate repair line, \myalgo applies its predicted repair (including concretization) and then compiles the resulting program. If the number of compilation errors in the program reduce, then this repair is accepted and the process is repeated for the remaining candidate repair lines.

\begin{table}[t]
	\centering
	\caption{{Comparison between \tracer and MACER on the single-line and multi-line test datasets. \myalgo achieves similar $\pred@\kac$ and repair accuracy as \tracer on the single-line dataset. On multi-line dataset, where programs require repairs on multiple different lines, \myalgo achieves 14\% improvement over \tracer.}}
	\label{tab:tra-mac-comp}
	\begin{tabular}{c|ccc|c}
			\toprule
			{Dataset} & \multicolumn{3}{c}{Single-line} \vrule & \multicolumn{1}{c}{Multi-line}\\\hline
			{Metric} & $\pred@1$ & $\pred@5$ & $\repl@5$ & $\repl@5$\\
			\midrule
			\tracer & 0.596 & 0.683 & 0.792 & 0.437\\
			\myalgo & 0.597 & 0.691 & 0.805 & \textbf{0.577}\\
			\bottomrule	
		\end{tabular}
\end{table}

\begin{table}[t]
	\centering
	\caption{{Comparison of all methods on the \deepfix dataset. Values take from $^{*}$\cite{GuptaPKS2017} and $^{\dagger}$\cite{GuptaKS2019}. \myalgo offers the highest repair accuracy on this dataset. The nearest competitor is \tracer that is 12.5\% behind. \myalgo offers a prediction time that is $4\times$ faster than \tracer and $2\times$ faster than the rest, and a train time that is $2\times$ faster than \tracer and more than $800\times$ faster than \rlassist.}}
	\label{tab:deepfix-data}
	\begin{tabular}{ccccc}
		\toprule
		{} & \deepfix & \rlassist & \tracer & \myalgo\\
		\midrule
		Repair Acc & 0.27$^{*}$ & 0.267$^{\dagger}$ & 0.439 & \textbf{0.566}\\
		Test Time & $<$1s$^{\dagger}$ & $<$1s$^{\dagger}$ & 1.66s & \textbf{0.45s}\\
		Train Time & - & 4 Days & 14 min & \textbf{7 min}\\
		\bottomrule
		\end{tabular}
\end{table}

\begin{figure}[t]%
	\centering
	\includegraphics[width=0.75\columnwidth]{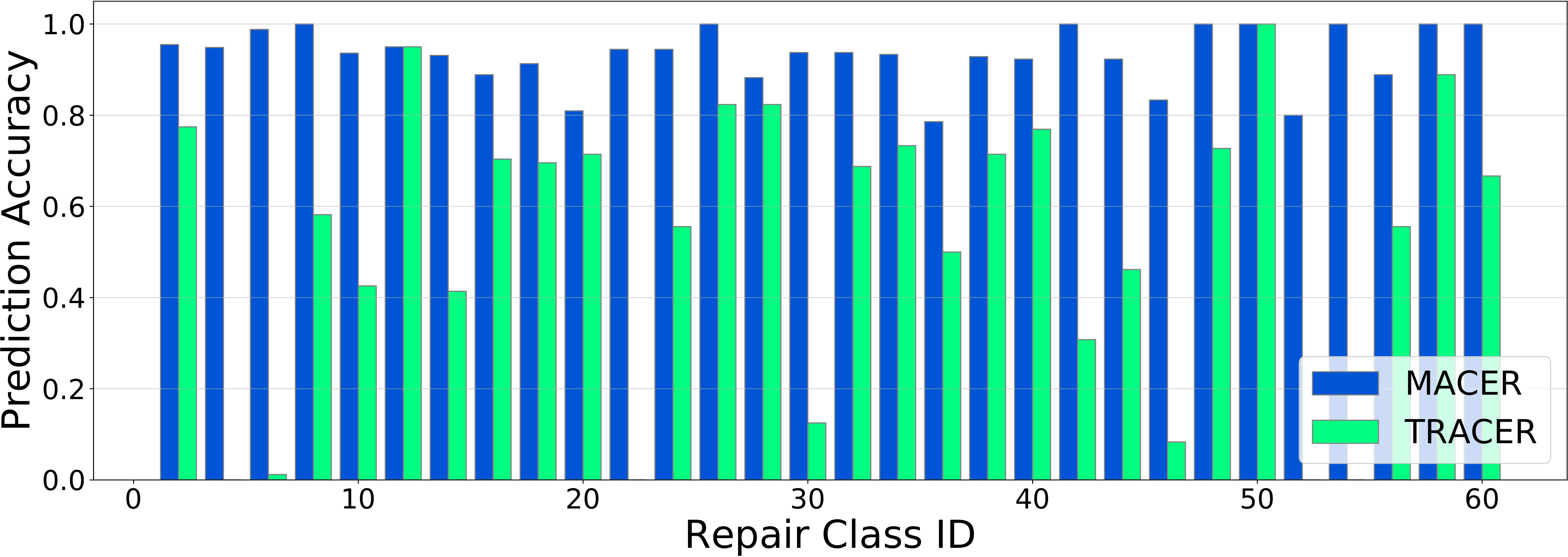}\\
	\includegraphics[width=0.75\columnwidth]{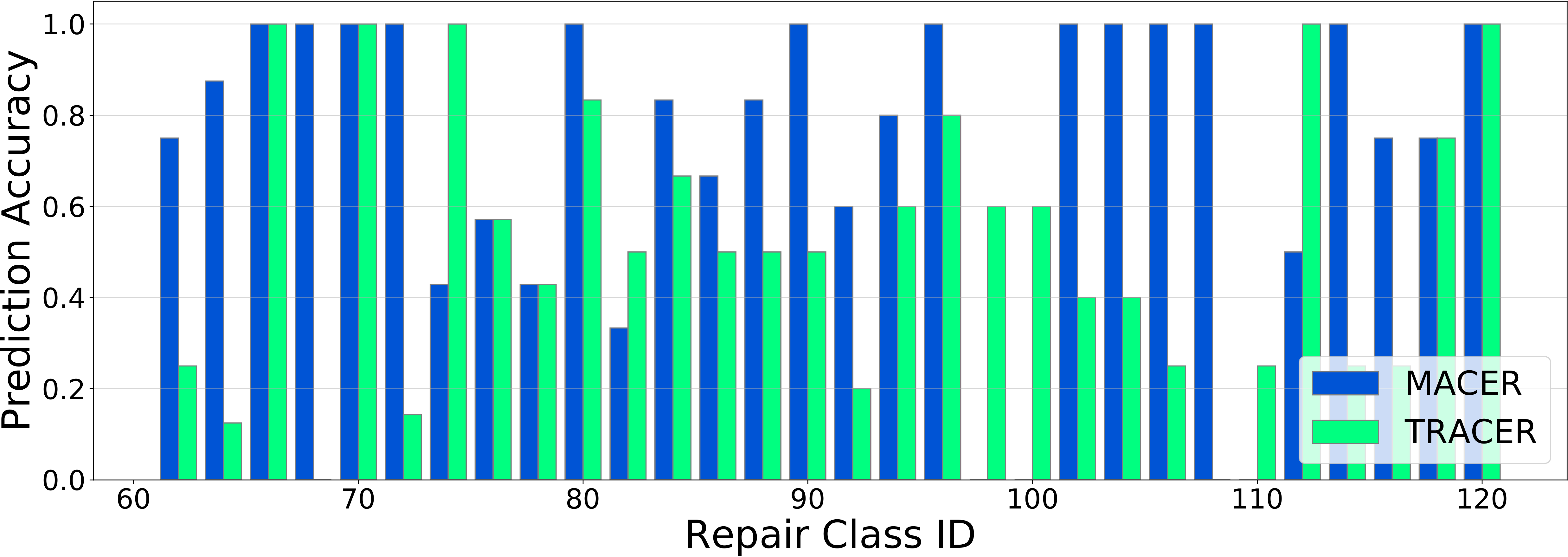}
	\caption{{The two figures compare \myalgo and \tracer on the head repair classes (top 60 in terms of popularity with 35+ training points in each class) and torso repair classes (top 60-120 with 15+ training points in each class). To avoid clutter, only 30 classes from each category are shown. \myalgo has a substantial lead over \tracer on head classes with an average of 20\% higher prediction hit rate (i.e. predicting the exact repair as desired by the student). On torso classes, \myalgo continues to dominate albeit with a smaller margin. On rare classes, the two methods are competitive.}}%
	\label{fig:head-torso}%
\end{figure}

\section{Experiments}
\label{sec:exps}

We compared \myalgo's performance against previous works, as well as performed ablation studies to study the relative contribution of its components. All \myalgo implementations\footnote{The \myalgo tool-chain is available at https://github.com/purushottamkar/macer/} were done using standard machine learning libraries such as \texttt{sklearn} \cite{sklearn} and \texttt{keras} \cite{keras}. Experiments were performed on a system with Intel(R) Core(TM) i7-4770 CPU @ 3.40GHz $\times$ 8 CPU having 32 GB RAM.

\subsection{Datasets}
We report \myalgo accuracy on three different datasets. All these datasets were curated from the same 2015-2016 fall semester course offering of CS-1 course at IIT-Kanpur (a large public university) where 400+ students attempted more than 40 different programming assignments. The dataset was recorded using Prutor~\cite{das2016prutor}, an online IDE. The \deepfix dataset\footnote{https://www.cse.iitk.ac.in/users/karkare/prutor/prutor-deepfix-09-12-2017.zip} contains 6,971 programs that fail to compile, each containing between 75 and 450 tokens \cite{GuptaPKS2017}. The \emph{single-line} (17,669 train program pairs + 4,578 test program pairs) and \emph{multi-line} (17,451 test program pairs) datasets\footnote{https://github.com/umairzahmed/tracer} released by \cite{AhmedKKKG2018} contain program pairs where error-repair is required, respectively, on a single line or multiple lines .\\

\subsection{Metrics}
We report our results on two metrics i) repair-accuracy, the popular metric widely adopted by repair tools, and ii) $\pred@\kac$, a metric introduced by \tracer~\cite{AhmedKKKG2018}. Repair accuracy denotes the the fraction of test programs that were successfully repaired by a tool i.e. all compilation errors were removed, thereby producing a correct program devoid of any compilation errors. On the other hand, $\pred@\kac$ metric captures the fraction of test programs where at least one of the top $\kac$ abstract repair suggestions (since \myalgo and other competing algorithms are capable of offering multiple suggestions for repair in a ranked list) exactly matched the student's own abstract repair.

The choice of the $\pred@\kac$ metric is motivated by the fact that the goal of program repair is not to generate any program that merely compiles. This is especially true of repair tools designed for pedagogical settings. rated by the tool is exactly same as the student generated one, at the abstraction level. The purpose of this metric is further motivated in the \S~\ref{app:kali}.

\subsection{Training Details}
\label{app:training}
Our training is divided into two parts, learning models to perform i) Repair Class prediction, and ii) Repair Location prediction. For repair class prediction we followed the prediction hierarchy shown in Figure ~\ref{fig:hierarchy}. The root node uses a feed forward neural net with two hidden layers of 128 nodes each. We tried \{1,2,3\} hidden layers with each layer containing nodes varying in \{128,256,512\}, and the structure currently used by us (with 2 hidden layers of 128 nodes each) was found to be the best.

For re-ranking the repair classes, we created prototype(s) of each class depending on the size of class using KMeans clustering. The second part of training is Repair Location Prediction. We followed one-vs-rest (OVR) approach and performed binary classification for each of the 1930 bigrams, each binary classification telling us whether the corresponding bigram is worthy of edits or not. We recall that these OVR classifiers were trained separately for all repair classes to allow greater flexibility. The binary classification was performed using standard implementations of decision trees that use Gini impurity to ensure node purity. Decision trees were chosen due to their speed of prediction and relatively high accuracy.

\subsection{A Naive Baseline and Importance of $\pred@\kac$}
\label{app:kali}
We consider a naive method \textit{Kali'} that simply deletes all lines where the compiler reported an error. This is inspired by Kali~\cite{qi2015analysis}, an erstwhile state-of-art semantic-repair tool that repaired programs by functionality deletion alone. This naive baseline \textit{Kali'} gets 48\% repair accuracy on the \deepfix dataset whereas \deepfix~\cite{GuptaPKS2017}, \tracer~\cite{AhmedKKKG2018} and \myalgo get respectively 27\%, 44\% and 56\% (Tab~\ref{tab:deepfix-data}). Although \textit{Kali'} seems to offer better repair accuracy than \tracer, its $\pred@1$ accuracy on the single-line dataset is just 4\%, compared to 59.6\% and 59.7\% by \tracer and \myalgo respectively (Tab~\ref{tab:tra-mac-comp}). This demonstrates the weakness of reporting on repair accuracy metric in isolation, and motivates the usage of additional complex metrics such as $\pred@\kac$, to better capture the efficacy of repair tools.

\begin{table*}[t]%
	\centering
    \caption{{Performance of \myalgo on several sample test instances. Pred? = Yes if \myalgo's top suggestion exactly matched the student's abstracted fix, else Pred? = No is recorded. Rep? = Yes if \myalgo's top suggestion removed all compilation errors else Rep? = Yes is recorded. ZS? records whether the example was a ``zero-shot'' test example where \myalgo had never seen the corresponding repair class in training data. On the first three examples, \myalgo not only offers successful compilation, but offers a repair that exactly matches that desired by the student. Note that the second example involves an undeclared identifier. In the next two examples, although \myalgo does not offer the exact match desired by the student, it nevertheless offers sane fixes that eliminate all compilation errors. In the fifth example, \myalgo errs on the side of caution and inserts a matching parenthesis rather than risk eliminating an unmatched parenthesis. The last two are zero-shot examples. Although \myalgo could handle one of the zero-shot cases gracefully, it could not handle the other case. Obtaining better performance on zero shot repair classes is a valuable piece of future work for \myalgo.}}
	\label{tab:egSrcTrgt}	
	\resizebox{\linewidth}!{
	\begin{tabular}{ll ll c}
		\hline
		\# & {Source-line} & {Target-line} & {\myalgo's Top Prediction} & (Pred?, Rep?, ZS?)\\ \hline
		1 & \texttt{scanf("\%c",\&a[i]\colorbox{red!30}{;}} & \texttt{scanf("\%c",\&a[i]);} & \texttt{scanf("\%c",\&a[i]\colorbox{green!30}{);}} & (Yes, Yes, No)\\
		
		2 & \texttt{for\colorbox{red!30}{(i}=0;i$<$n;i++)} & \texttt{for(int i=0;i$<$n;i++)} & \texttt{for\colorbox{green!30}{\!(int i\!}=0;i$<$n;i++)} & (Yes, Yes, No)\\
		
		3 & \texttt{if(x==y)printf("Y");\colorbox{red!30}{break;}} & \texttt{if(x==y)printf("Y");} & \texttt{if(x==y)printf("Y")\colorbox{green!30}{;}} & (Yes, Yes, No)\\
		
		4 & \texttt{for(i=0;\colorbox{red!30}{i$=<$N};i++)} & \texttt{for(i=0;i$<=$N;i++)} & \texttt{for(i=0;\colorbox{green!30}{i$<$N};i++)} & (No, Yes, No)\\
		
		5 & \texttt{if\colorbox{red!30}{\!(\!}(a[j]==' ')} & \texttt{if(a[j]==' ')} & \texttt{if((a[j]==' ')\colorbox{green!30}{\!)\!}} & (No, Yes, No)\\
		6 & \texttt{int n;\colorbox{red!30}{n=q;}}& \texttt{int n;}& \texttt{int n;\colorbox{green!30}{n=0;}}& (No, Yes, Yes)\\
		
		7 & \texttt{c=sqrt(\colorbox{red!30}{a\^{}2+b\^{}2});} & \texttt{c=sqrt(a*a+b*b);} & \texttt{c=sqrt(\colorbox{red!30}{a\^{}2+b\^{}2});} & (No, No, Yes)\\
        \bottomrule
	\end{tabular}
    }
\end{table*}

\begin{figure}[t]%
	\centering
	\includegraphics[width=0.75\columnwidth]{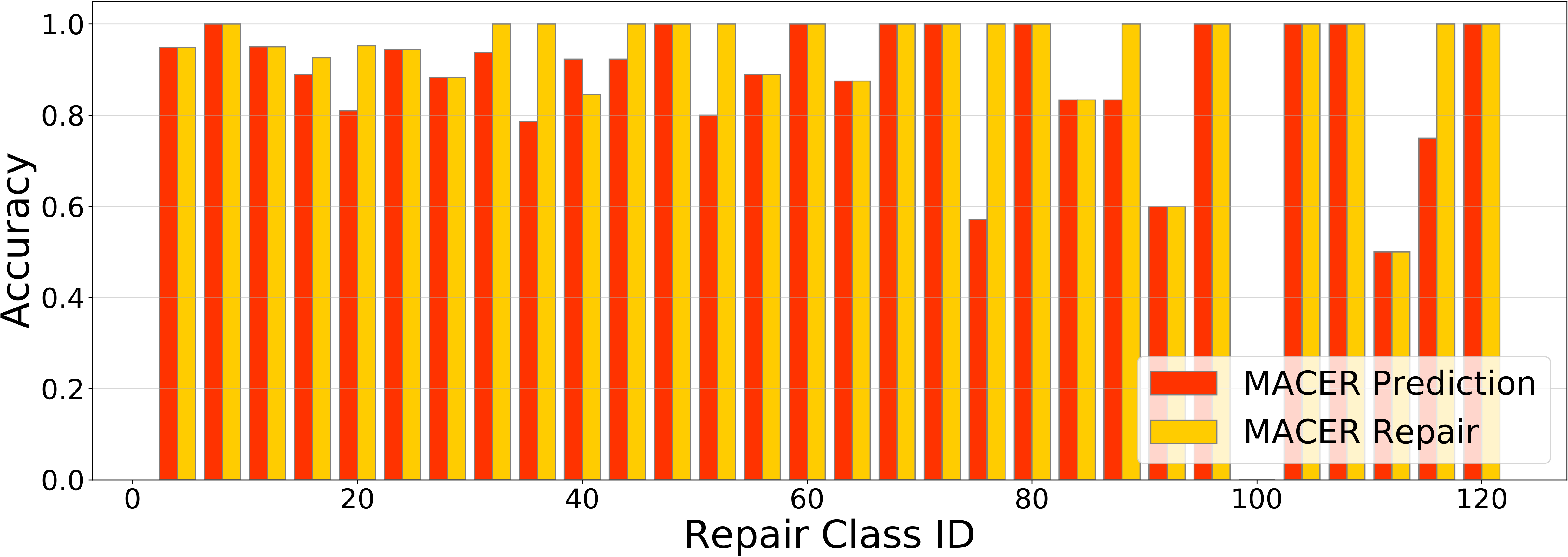}
	\caption{{A graph comparing the prediction and repair hit rate of \myalgo on the top 120 most popular repair classes. To avoid clutter, every 4$\nth$ class is shown. For these popular classes -- which still may have as low as 15 training data points -- \myalgo frequently achieves perfect or near perfect score in terms of prediction accuracy or repair accuracy or both.}}%
	\label{fig:pvr}%
\end{figure}

\begin{figure}[t]%
	\centering
	\includegraphics[width=0.3\columnwidth]{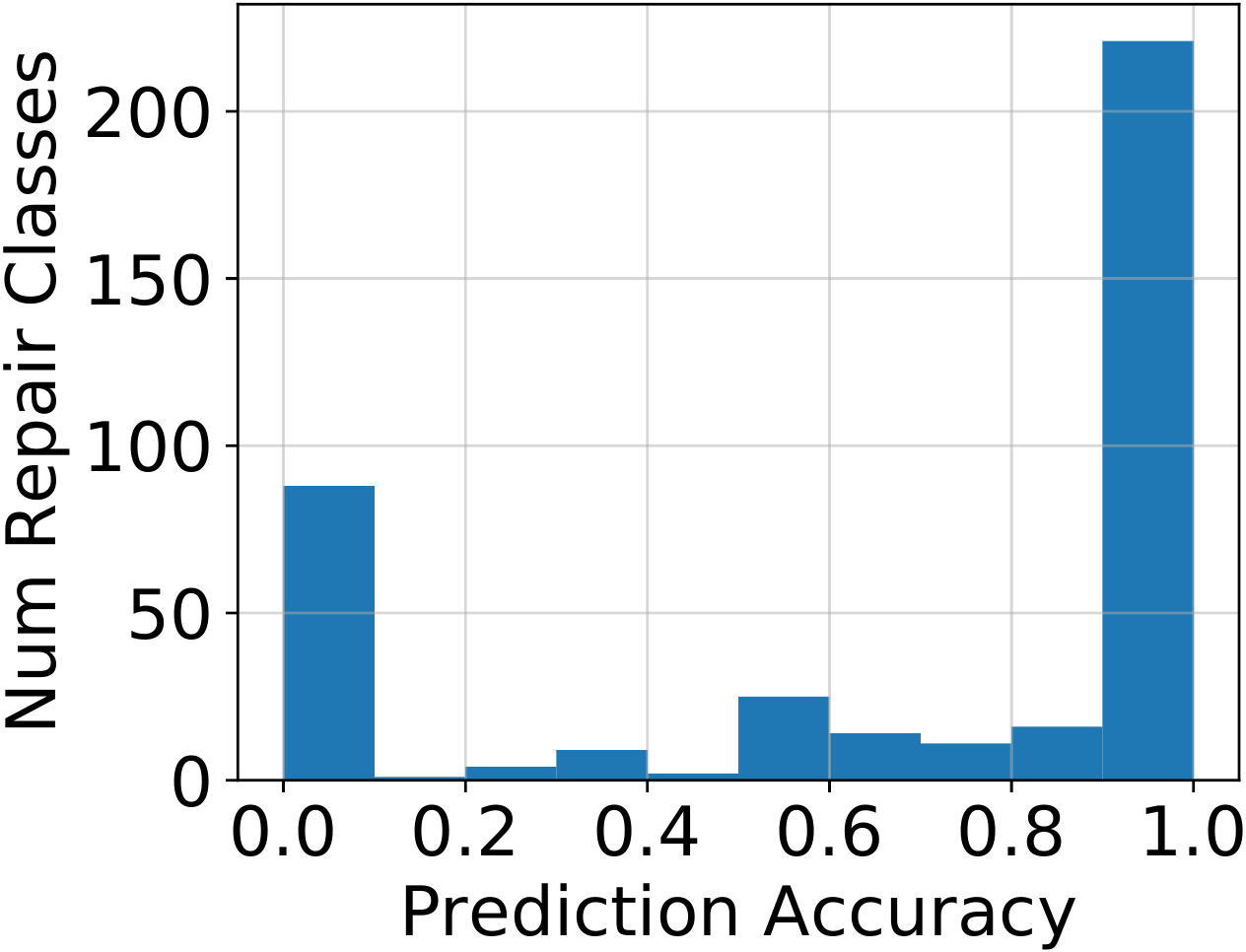}%
	\includegraphics[width=0.3\columnwidth]{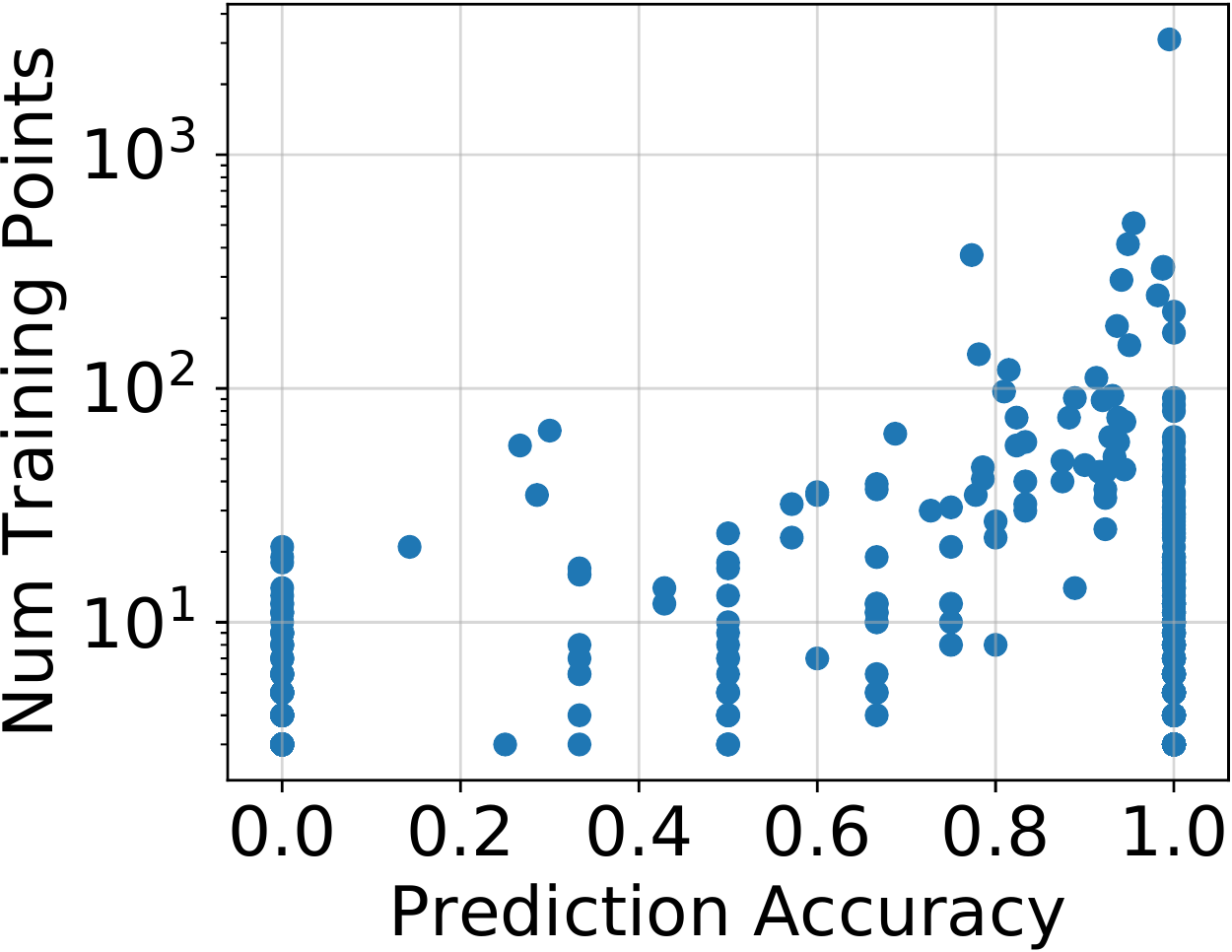}%
	\caption{A study similar to the one presented in Fig~\ref{fig:repair} but with respect to prediction (exact match) accuracy instead of repair accuracy. A study of the prediction accuracy offered by \myalgo on repair classes with at least 3 training data points -- a total of 391 such classes were there. On a majority of these classes 221/391 = 56\%, \myalgo offers greater than 90\% prediction accuracy. On a much bigger majority 287/391 = 73\% of these classes, \myalgo offers more than 50\% prediction accuracy. The second graph indicates that \myalgo's prediction accuracy drops below 50\% only on classes which have less than around 30 points.}%
	\label{fig:pred}%
\end{figure} 

\begin{figure}[t]%
	\centering
	\includegraphics[width=0.3\columnwidth]{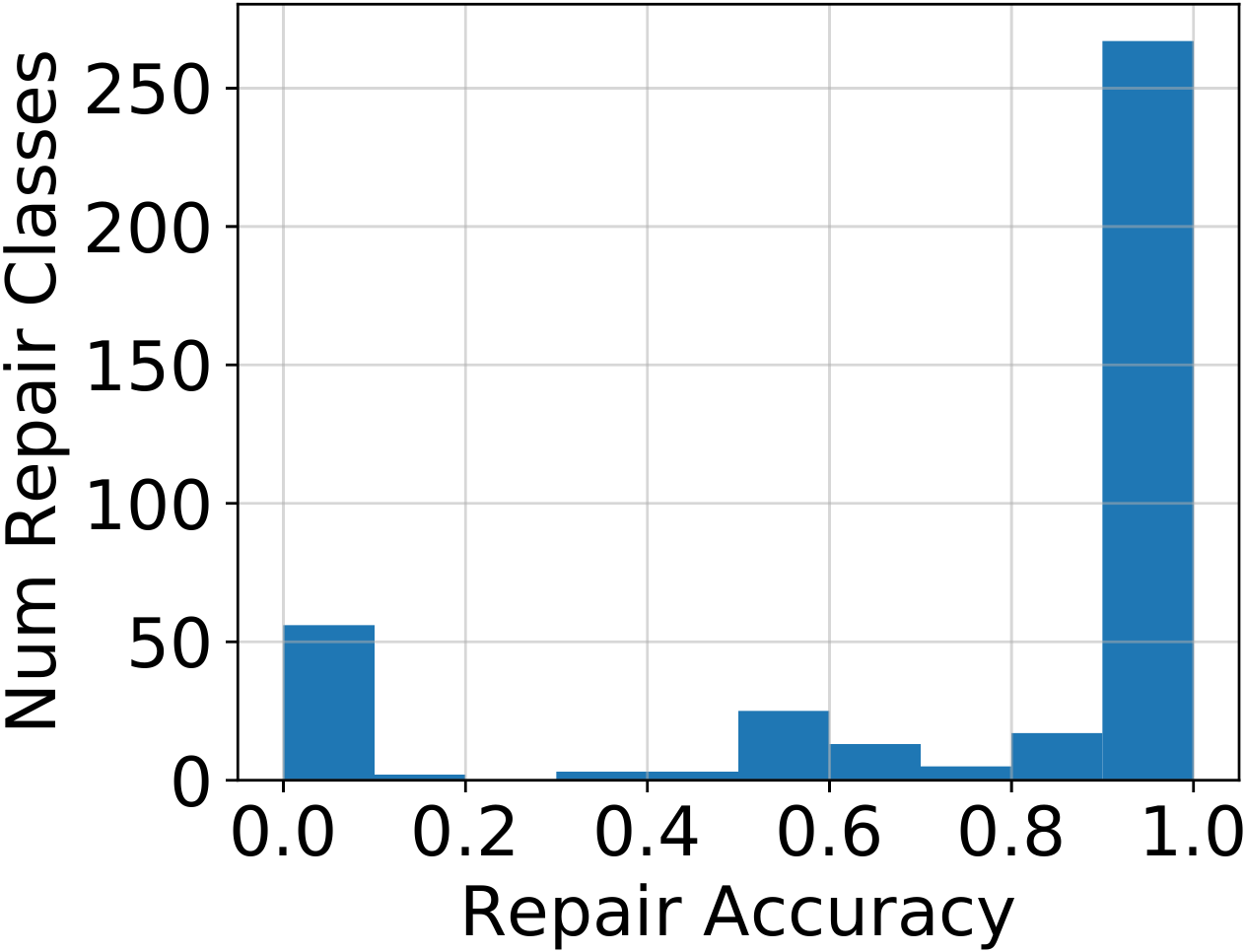}%
	\includegraphics[width=0.3\columnwidth]{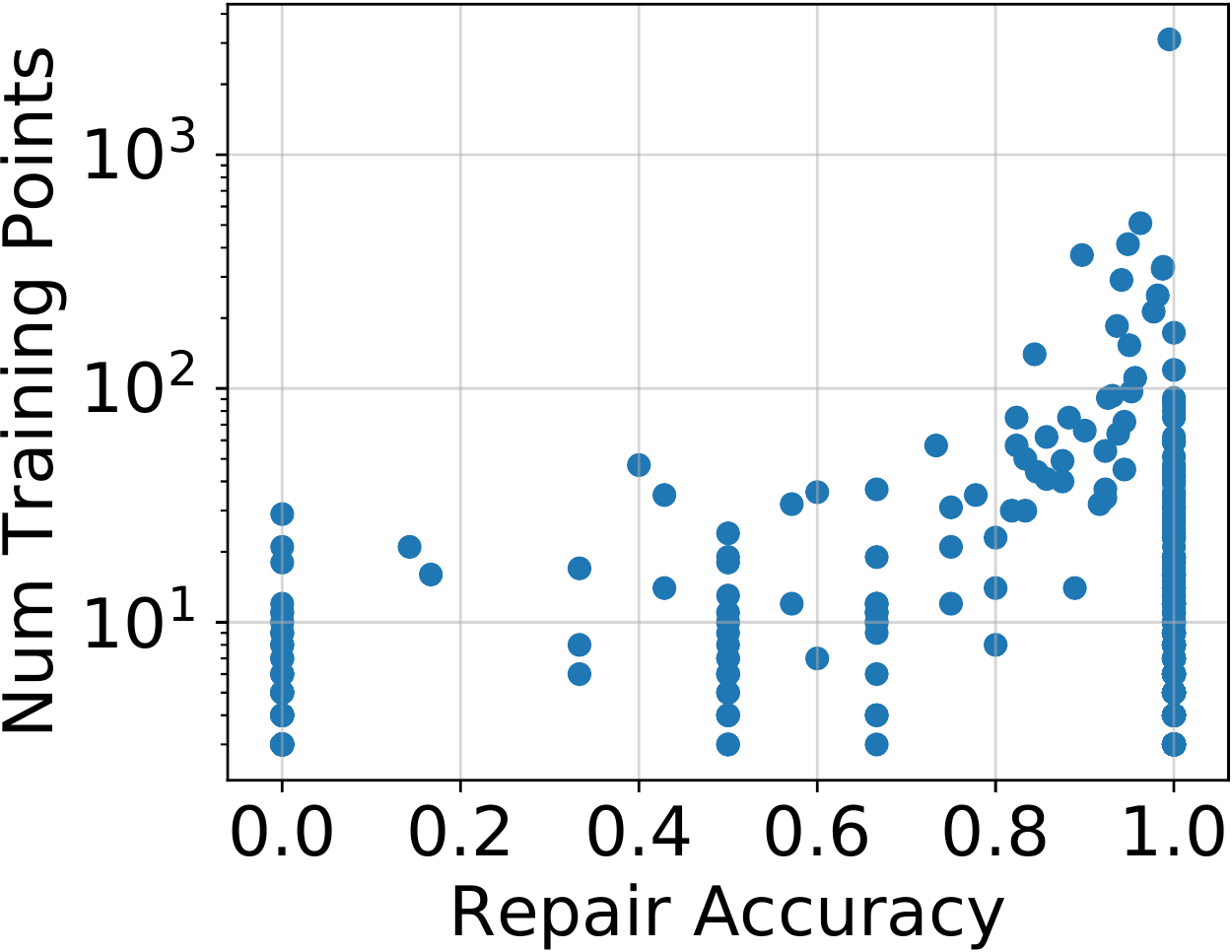}%
	\caption{{A study of the repair accuracy offered by \myalgo on repair classes with at least 3 training data points -- a total of 391 such classes were there. On a majority of these classes 267/391 = 68\%, \myalgo offers greater than 90\% repair accuracy. On a much bigger majority 327/391 = 84\% of these classes, \myalgo offers more than 50\% repair accuracy. The second graph indicates that \myalgo's repair accuracy drops below 50\% only on classes which have less than around 30 points. This indicates that \myalgo is very effective at utilizing even small amounts of training data.}}%
	\label{fig:repair}%
\end{figure}

\begin{table}
	\centering
	\caption{{An ablation study on the differential contributions of \myalgo's components. ZS stands for ``zero-shot''. For the ``ZS included'' column all test points are considered while reporting accuracies. For the ``ZS excluded'' column, only those test points are considered whose repair class was observed at least once in the training data. RR stands for Reranking. RCP stands for Repair Class Prediction, RLP stands for Repair Location Prediction. RCP = P (resp. RLP = P) implies that we used the repair class (resp. repair location) predicted by \myalgo. RCP = G (resp. RLP = G) implies that we used the true repair class (resp. true repair profile vector). It is evident from the difference in the results of the first two rows that (whether we include ZS or not), reranking gives 10-12\% boost in both prediction accuracy. This highlights the importance of reranking in the presence of rare classes. Similarly, it can be seen that predicting the repair class (resp location) correctly accounts for 5-12\% (resp. 6\%) of the performance. The final row shows that \myalgo loses 6-8\% performance owing to improper repair application/concretization. In the last two rows, $\pred@1$ is higher than $\repl@1$ (1-2\% cases) owing to concretization failures -- even though the predicted repair matched the student's repair in abstracted form, the program failed to compile after abstraction was removed.}}
	\label{tab:ablat-study}
	\begin{tabular}{lllllll}
	\toprule
	{} & {} & {} & \multicolumn{2}{c}{ZS included} & \multicolumn{2}{c}{ZS excluded}\\
	\midrule
	RR & RCP & RLP & \pred@1 & \repl@1 & \pred@1 & \repl@1\\
	\hline
	OFF & P & P & 0.492 & 0.599 & 0.631 & 0.706\\
	ON & P & P & 0.597 & 0.703 & 0.757 & 0.825\\
	ON & G & P & - & - & 0.885 & 0.877\\
	ON & G & G & - & - & 0.943 & 0.926\\
	\bottomrule
	\end{tabular}
\end{table}

\subsection{Breakup of Training Time}
Of the total 7 minute train time (see Tab~\ref{tab:deepfix-data}), \myalgo took less than 5 seconds to create repair classes and repair profiles from the raw dataset. The rest of the training time was taken up more or less evenly by repair class prediction training (tree ranking + reranking) and repair profile prediction training.

\subsection{Comparisons with other methods}
The values for $\pred@k$ (resp. $\repl@k$) were obtained by considering the top $k$ repairs suggested by a method and declaring success if any one of them matched the student repair (resp. removed compilation errors). For $\pred@\kac$ computations,  all methods were given the true repair line and did not have to perform repair line localization. For $\repl@\kac$ computations, all methods had to localize then repair. Tabs~\ref{tab:tra-mac-comp} and \ref{tab:deepfix-data} compare \myalgo with competitor methods. \myalgo offers superior repair performance at much lesser training and prediction costs. Fig~\ref{fig:head-torso} shows that \myalgo outperforms \tracer by $\approx$ 20\% on popular classes while being competitive or better on others.

\subsection{Ablation studies with \myalgo}
To better understand the strengths and limitations of \myalgo, we report on further experiments. Fig~\ref{fig:pvr} shows that for top 120 most popular repair classes (which still may have as low as 15 training data points), \myalgo frequently achieves perfect or near perfect score in terms of prediction accuracy or repair accuracy or both. Figs~\ref{fig:pred} and \ref{fig:repair} shows that \myalgo is effective at utilizing even small amounts of training data and that its prediction accuracy drops below 50\% only on repair classes which have less than 30 examples in the training set. Tab~\ref{tab:egSrcTrgt} offers examples of actual repairs by \myalgo. Although it performs favorably on repair classes seen during training, it often fails on \emph{zero-shot} repair classes which were never seen during training. Tab~\ref{tab:ablat-study} presents an explicit ablation study analyzing the differential contributions of \myalgo's individual components on the single-line dataset. Re-ranking gives 10-12\% boost to both $\pred@\kac$ and repair accuracy. Predicting the repair class (resp. profile) correctly accounts for 5-12\% (resp. 6\%) of the performance. \myalgo loses a mere 6\% accuracy on account of improper repair application. For all figures and tables, details are provided in the captions.

\clearpage

\section{Conclusion}
\label{sec:concl}
In this paper we presented \myalgo, a novel technique to accelerated compilation error-repair. A key contribution of \myalgo is a fine-grained segregation of the error repair process into efficiently solvable ranking and labelling problems. These reductions are novel in this problem area where most existing techniques prefer to directly apply a single powerful generative learning technique instead. \myalgo offers significant advantages over existing techniques namely superior error repair accuracy on various error classes and increased training and prediction speed. Targeting rare error classes and ``zero-shot'' cases (Tab~\ref{tab:egSrcTrgt}) is an important area of future improvement. A recent large scale user-study~\cite{seet2020} demonstrated that students who received automated repair feedback from \tracer~\cite{AhmedKKKG2018} resolved their compilation errors faster on average, as opposed to human tutored students; with the performance gain increasing with error complexity. We plan to conduct a similar systematic user study in the future, to better understand the correlation between the $\pred@\kac$ metric scores and error resolution efficiency (performance) of students.

\section*{Acknowledgments}
The authors thank the reviewers for helpful comments and are grateful to Pawan Kumar for support with benchmarking experiments. P. K. thanks Microsoft Research India and Tower Research for research grants.

\clearpage

\bibliographystyle{plainurl}
\bibliography{refs}

\appendix
\normalsize

\section{Hierarchical Repair Class Ranking}
\label{app:repair-class-rank-app}

Let $T$ denote the tree in Fig~\ref{fig:hierarchy} with root $r(T)$. The leaves of $T$ correspond to individual repair classes. The set of leaf nodes of a subtree rooted at any node $t$ will be denoted by $L(t)$. The set of children of a node $t$ will be denoted by $C(t)$ and the parent of node $t$ will be denoted by $P(t)$.

The set of nodes on the path from the root $r(T)$ to any leaf $l \in L(r(T))$ is denoted by $W(l)$. Note that each leaf $l$ corresponds to a repair class $c_l \in [1016]$. For any node $t$, let the indicator random variable $V_t$ indicate if we visited node $t$. Then using the chain rule, we can express
\[
\P{y = c \cond \vx} = \P{V_{l_c} = 1 \cond \vx} = \prod_{t \in W(l_c)}\P{V_t = 1 \cond \vx, V_{P(t)} = 1}
\]
The correctness of the above can be deduced from the fact that $V_t = 1$ implies $V_{P(t)} = 1$. Given that all nodes in our tree train their classifiers probabilistically using the cross entropy loss, we are readily able to, for every internal node $tn$ and its, say $k$ children $c_1,\ldots,c_k$, assign the probability $\P{V_{c_i} = 1 \cond \vx, V_t = 1}$ using the sigmoidal activation (for binary split at the root), or the softmax activation (for all other multi-way splits). This allows us to compute the score $\hat s_c(\vx) := \P{y = c \cond \vx}$ for any repair class $c$ by just traversing the tree from the root node to the leaf node corresponding to the repair class $c$.

\end{document}